# Security Strength Indicator in Fallback Authentication: Nudging Users for Better Answers in Secret Questions

Awanthika Senarath
*Australian Centre for Cyber Security, University of New South Wales (UNSW Canberra), the Australian Defense Force Academy*

Nalin A. G. Arachchilage
*Australian Centre for Cyber Security, University of New South Wales (UNSW Canberra) the Australian Defense Force Academy*

B. B. Gupta
*National Institute of Technology, Kurukshetra, Haryana, India*

## Abstract

*In this paper, we describe ongoing work that focuses on improving the strength of the answers to security questions. The ultimate goal of the proposed research is to evaluate the possibility of nudging users towards strong answers for ubiquitous security questions. In this research we are proposing a user interface design for fallback authentication to encourage users to design stronger answers. The proposed design involves visual feedback to the user based on mnemonics which attempts to give visual feedback to the user on the strength of the answer provided and guide the user to creatively design a stronger answer.*

## 1. Introduction

Fallback authentication is the backup, which approach a user takes to retrieve his/her account in case of loss of password [4]. Almost all sites, Google, Microsoft, and Facebook that involve user accounts have incorporated fallback authentication methods to facilitate users to regain their accounts in case user lose the password. Fallback authentication enables account regaining possible without forcing the user to go through customer assistance email/phone calls which was the case in the early stages of account recovery [3].

Mechanisms of providing security in web are always a trade-off between security and usability [23]. This research evaluates the possibility of improving the security aspect of fallback authentication whilst keeping the usability aspect intact.

Web based accounts are inevitably vulnerable to identity theft and a fallback authentication approach is required for convenient account recovery [3]. Fallback authentication mainly follows three common approaches, SMS based account recovery [10], verification email based account recovery and security question based account recovery [9]. Account recovery using a mobile device is attractive as it provides additional mobility aspect to the user. The most remarkable feature of the mobile environment is mobility of the device, mobility of the service as well as mobility of the user [16]. However, this mobility feature itself could oppose security; for example, SMS messages may fail if the user does not have access to their mobile telephones while travelling overseas. On the other hand, mobile telephones are not only prone to get stolen and lost but also frequently shared among family and peers. Both the first two methods lack usability aspect as it requires the use of a secondary account, personal mobile phone or personal email account for successful account recovery. In the case where the user loses access to all his/her personal information and lose access to his/her accounts this approach may not be a viable solution, as it depends on a secondary account to recover the primary account.

The other approach utilized for fallback authentication is the security questions based account recovery mechanism [3]. It is commonly argued that fallback authentication should always involve personal information for account recovery, which need not to be memorized.

This is on the basis that the answers given to security questions should be more memorable than the password [5] [6] [7]. Common examples being "mother's maiden name", "your favorite movie in childhood". This type of questions are selected mainly due to the understanding of the user approaching the fallback authentication to reclaim their account when they have lost the original credentials. Therefore, the recovery approach needs to be knowledge based and not memory based [21]. These are called cognitive passwords. A cognitive password is a form of knowledge-based authentication that requires a user to answer a question, presumably something they intrinsically know, to verify their identity. Cognitive passwords have high memorability rates, but they are commonly used and are easily guessable [21].

Security experts and phishing attackers are in a rat race today. On the one hand, security experts with the help of software designers and developers will continue to improve phishing and spam detection tools. Nevertheless, people are the weakest link in information security [11, 14, 25]. On the other hand, attackers will not hesitant to learn new techniques and change their strategies according to human defects to make a phishing attack successful [24, 25]. To prevent this, phishing education needs to be considered [6, 15, 24, 25, 35, 36, 38, 39 and 52].

## 2. Related Work

In 1990, Zviran and Haga's study showed that cognitive passwords, while being easily memorable are hard to guess, even for close friends [13]. Yet, having a method of authenticating purely based on one's personal knowledge may reduce the security aspect, though it improves the usability. Best example being Republican vice presidential candidate Sarah Palin's Yahoo! Email account being "hijacked" in the run-up to the 2008 US election [1]. The "hacker" simply used the password reset prompt and answered her security questions [2]. It is evident that the hacker in this case did not need any password hacking methods, other than mere social engineering knowledge to find out the victim's personal information by searching their personal profile.

Podd et al. [25] conducted a questionnaire survey study measuring participants' recall and guessing rates of conventional, cognitive and word association passwords. The study employed 86 Massey University undergraduates. Participants completed a questionnaire covering all three password types, as well as returned after two weeks for a recall test. Each participant was also asked to nominate a "significant other" (e.g. parent, partner, etc.), who attempted to guess the participant's answers. The study findings revealed that on average, cognitive items produced the highest recall rates (80%), however, the guessing rate was also significantly high (39.5%). Word associations produced significantly low guessing rates (7%), however, it has been shown that response words were poorly recollected (39%). Nevertheless, authors concluded that both cognitive items and word associations have a sufficient promise as password techniques (in this case recall and guessing) to warrant further investigation.

Security aspect of Fallback authentication has been much researched upon [21]. Even though extensive research has been conducted in improving the security of password authentication [4], [5], [6], [7], and also in introducing novel techniques for fallback authentication [14], [21], adequate attention has not been paid to improve the usability and security aspects of the most commonly used fallback authentication technique for many web based accounts [4] [21] which is Security questions based fallback authentication process.

The capability to verify the user identity when an account hijacking attempt has occurred is an integral part of the login risk analysis system [12]. Google researchers along with academics have revealed that current security questions are neither secure nor reliable enough to be used as a backup mechanism to reclaim a lost account [3]. Their argument was that security questions suffer from a fundamental flaw of usable security: the security questions and their answers are either somewhat secure or usable, but rarely both. They also stressed that security questions can still be useful when the risk level is considered low [3]. To design a better extra level of security, it is worth understanding the strength of the answers users provided for security questions.

This paper attempts to introduce a set of guidelines to design an interface called "secret question meter". The "secret question meter" interface provides visual feedback on the strength of answers given in security questions to nudge users towards stronger answers. The designed visual feedback to the user based on mnemonics attempted to nudge the user towards strong answers for their chosen security questions. The visual representation of answers' strength to security questions is often presented as a colored bar on screen. Furthermore, our "secret question meter" interface provides suggestions to assist users in selecting strong security questions and their answers.

We have integrated suggestions based on mnemonics for stronger answers if the user decides to go along with the weak answer irrespective of the feedback from the strength indicator. Mnemonics have been tested and proven to improve memorability on stronger passwords rather than random passwords with digits and special characters to improve strength [15]. All these mechanisms are not enforced on the user and work as suggestions where the user holds the final decision to set the answer they feel best.

## 3. Proposed Design Guidelines

We are proposing an interface design to walk the user through a process of setting up a stronger fallback authentication layer with an extra level of security.

The proposed design involves five security questions, three pre-defined and selectable from a drop down list of questions, and two defined by the user. User selected security questions are proven to be more effective in fallback authentication configuration as user defined questions are harder to guess than the predefined questions that are ubiquitous. Google allows users to create their own security questions as a part of setting up their fallback authentication method [3]. Yet, depending solely on user created security questions may also introduce lower security level as users may select simpler questions [3].

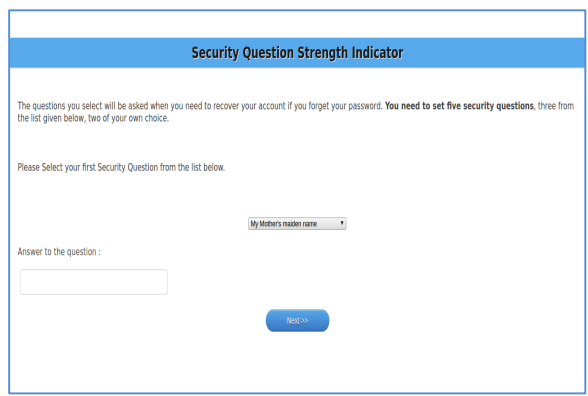

Figure. 1. Security Question Interface.

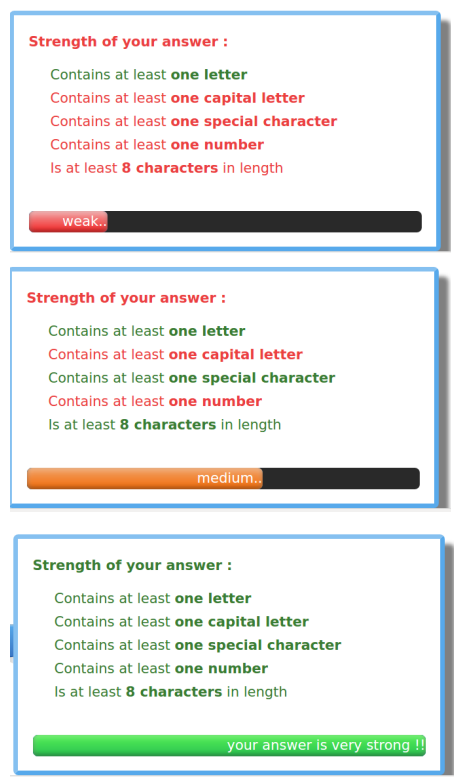

Figure. 2. Strength Indicator Meter. (a) weak (b) medium strength (c) Strong

On answering the questions, the proposed design provides the users with a feedback dialog which would give the user an idea of the strength of the answers they should provide, to have a secure answer adhering to five guidelines. According to Blase et al. [17] password strength meters with a variety of visual appearances lead users to create longer passwords. However, significant increases in resistance to a password-cracking algorithm were only achieved using meters that scored passwords that leads users to include more digits, symbols, and uppercase letters into their password. The proposed design involves a meter similar to the password strength meter to evaluate the strength of the answers given by the user. Many popular websites, from Google to Twitter, employ password meters [18]. These five guidelines are similar to the common parameters used in password strength indicators [17] which are proven to be effective. The design has an animated interface to enhance the visual appearance. It will dynamically display the strength of the answer the user type and will motivate the user towards a stronger, more secure answer [19]. The strength of the answer is measured by the following parameters.

- Has at least one capital letter
- Has at least one digit
- Has at least one special character
- Has at least one letter
- At least eight characters long

These animated interfaces would give the user an idea of the strength of the answer they have already set-up. It not only indicates the strength level of the answer provided with the parameters being displayed in the meter, but also contains a color bar that goes from red (weak), orange (strong) to green (very strong) to effectively indicate the strength of the word typed [19]. Yet, these guidelines and strength indicators are designed for password (which could be any random text) setup process, and may not suite the security question answers (which has a fixed answer). To adhere to the given guidelines users need to creatively design their answers.

Mnemonics is an approach that could be used to generate strong yet memorable text phrases. Primal et al [15] proposes mnemonics to be used as a solution for memorable stronger passwords. Mnemonics uses personal memories and use them in their passwords/answers that allow the individual to more easily and efficiently recall certain personal experiences [20]. We have incorporated mnemonics to give hints to the user to creatively set up stronger answers that adheres to the parameters specified by the strength indicator meter via simple dialogs.

If the user decides to proceed with a weak answer, i.e. the user neglects the strength indicator feedback and save a weak answer for a particular security question, we display a mnemonic based suggestion to the user to give him an idea how to set a stronger answer for a security question. This is similar to the passive aggressive password meter design by Tim Holman and Tobias van Schneider [24]. This approach is passive yet effective in coaxing the user towards a better answer. We have integrated the same concept to incorporate mnemonics in designing a stronger answer for the security questions. The dialog is phrased with simple text that encourages user towards a stronger answer for their security question which is both personal knowledge and not easily guessable.

The dialog box provides sample answers for questions in the drop down list of the question selection interface. The suggestions are simple and straightforward, for example,
  **Question:** What is your favorite sport?
  **Mnemonic based answer:** CrickICC15@Aus.

The dialog also carries an explanation educating the user how the mnemonic answer was composed. In the example provided the explanation would be, "My favorite sport is cricket, my favorite cricket team is Australia and they won the ICC world cup in 2015". Primal et al [15] emphasizes that mnemonic based words are both memorable and strong when combined with personal memories which has high recall rate.

## 4. Conclusion and Future Work

With this approach we hope to improve the current approach used in security question fallback authentication systems. As previously mentioned hacking an account by answering the security questions of the fallback authentication process does not require any special skill other than simple social engineering knowledge. Hence, the proposed design attempts to ensure that the answers given to security questions are not easily guessable for a third party individual to gain easy access to the user account. The proposed interface design is expected to improve the attention paid by user on the fallback authentication security question setup process. It educates the user on mnemonics based answer designing for better security with usability.

We hope to conduct user survey with the use of the proposed interface to evaluate the impact of the interactive strength indicator and answer suggestion dialogs on the end user. Also we hope to improve the design depending on the feedback of the survey participants.